\documentclass[fleqn,twoside]{article}
\usepackage{espcrc2}
\usepackage{amssymb}
\usepackage[figuresright]{rotating}
\newenvironment{Eqnarray}%
          {\arraycolsep 0.14em\begin{eqnarray}}{\end{eqnarray}}

\newcommand{\bc}{\begin{center}}
\newcommand{\ec}{\end{center}}
\newcommand{\eq}{\begin{equation}}
\newcommand{\ee}{\end{equation}}
\newcommand{\ea}{\begin{Eqnarray}}
\newcommand{\eea}{\end{Eqnarray}}
\newcommand{\be}{\begin{equation}}
\newcommand{\bea}{\begin{eqnarray}}

\newcommand{\AmS}{{\protect\the\textfont2
  A\kern-.1667em\lower.5ex\hbox{M}\kern-.125emS}}

\title{
\thispagestyle{empty}
\vspace{-25mm}
\rightline{\small DESY 03-117~~~~~}
\rightline{\small ITEP-LAT/2003-24~~~~~}
\rightline{\small KANAZAWA-03-24~~~~~}
\vspace{10mm}
Finite temperature phase transition in full QCD with $N_f=2$ flavors of
clover fermions at $N_t=8$ and 10
\thanks{Talk given by Y.~N. at Lattice'03.}
}
\author{Y. Nakamura\address{Kanazawa University, Kanazawa 920-1192, Japan\\[-0.5em]},
V. Bornyakov$^{\rm a}$\address{Institute for High Energy Physics, RU-142284 Protvino, Russia\\[-0.5em]},
M.N. Chernodub $^{\rm a}$\address{ITEP, B.Cheremushkinskaya 25, RU-117259 Moscow, Russia\\[-0.5em]},
Y. Koma$^{\rm a}$ ,
Y. Mori$^{\rm a}$ ,
M.I. Polikarpov$^{\rm c}$, \\
G. Schierholz\address{NIC/DESY Zeuthen, Platanenallee 6, D-15738 Zeuthen, Germany\\[-0.5em]} ,
A. Slavnov\address{Steklov Mathematical Institute, Vavilova 42, RU-117333 Moscow, Russia\\[-0.5em]} ,
H. St\"uben\address{Konrad-Zuse-Zentrum f\"ur Informationstechnik Berlin, D-14195 Berlin, Germany\\[-0.5em]},
T. Suzuki$^{\rm a}$,
P. Uvarov$^{\rm c}$, and
A.I. Veselov$^{\rm c}$ \\
}

\begin{document}
\begin{abstract}
We present results for QCD with $N_f=2$ flavors of dynamical quarks using
nonperturbatively improved
Wilson fermions at finite temperature on $16^3 \times 8$ and $24^3 \times 10$
lattices. We determine the transition temperature in the range
of quark masses $0.6 \lesssim m_\pi/m_\rho \leq 0.8$.
After fixing the Maximal Abelian gauge we
investigate the contribution of Abelian monopoles to the Polyakov loop,
Polyakov loop susceptibility and confirm Abelian and monopole dominance in
full QCD.
\end{abstract}
\maketitle

\section{INTRODUCTION}
\vspace{-1mm}
In order to obtain predictions for the real world from lattice QCD, we have to
extrapolate the lattice data to the continuum and to the chiral limits.
Recently the Bielefeld group~\cite{kpe} and the CP-PACS
collaboration~\cite{aakcp} using different fermion actions obtained consistent
values for the critical temperature $T_c$ in the chiral limit, 
albeit on rather coarse lattices at $N_t=4$
and 6. Edwards and Heller~\cite{eh} determined $T_c$ for $N_t=4$, 6 using
nonperturbatively improved Wilson fermions. We compute $T_c$ on finer
lattices with $N_t=8$ and 10 with high statistics. Our results for $N_t=8$ 
were reported in Ref.~\cite{previous}.

\section{SIMULATION}
\vspace{-1mm}
We use fermionic action for nonperturbatively $O(a)$ improved Wilson fermions:
\ea
S_F &=&  S^o_F - \frac{\rm i}{2} \kappa\, g\,
c_{sw} a^5
\sum\nolimits_x \bar{\psi}(x)\sigma_{\mu\nu}F_{\mu\nu}\psi(x) \nonumber,
\eea
where $S^o_F$ is the original Wilson action, $c_{sw}$ was calculated in
\cite{Jansen:1998mx}.

\begin{figure}[!h]
\centerline{\includegraphics[angle=0,scale=0.28,clip=true]{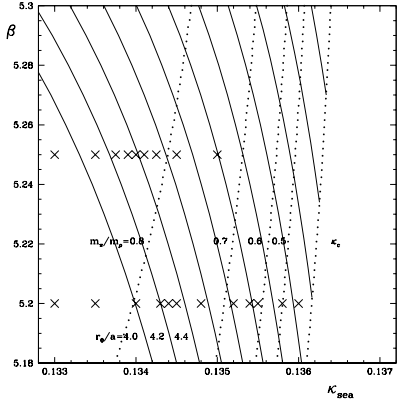}}
\vskip -11mm \caption{\it The lines of constant $r_0 /a$ and $m_\pi /m_\rho$ at
$T=0$. Crosses correspond to parameters used in this work.} 
\label{constant_DIK.ps}
\vskip -6mm
\end{figure}

Configurations are generated on $16^3 \times 8$ ($\beta=5.2$ and $5.25$) and
$24^3 \times 10$ ($\beta=5.2$) lattices at various $\kappa$.
The values of $\kappa$ and corresponding number of trajectories for
$16^3 \times 8$ and $24^3 \times 10$ lattices can be found in Ref.~\cite{previous} 
and Table 1, respectively.
The number of configurations for $24^3 \times 10$ lattice is not large
enough and results for this lattice are preliminary.
\vskip -1.5mm
{\footnotesize
\bc
\begin{tabular}{|c|c|c|c|c|} \hline
$\kappa$ & $0.1352$ & $0.1354$ & $0.1356$ & $0.1358$ \\ \hline
\# traj. & $ 5,400$ & $ 7,400$ & $3,130 $ & $ 1,650$ \\  \hline
\end{tabular}\\
\vskip 0.5mm
{\normalsize{Table 1:{\it Simulation statistics on $24^3 \times 10$.}}}
\ec
}
\vskip -2mm
We use results obtained at T=0 to fix the scale. An updated version
of the contour plot of lines of constant $r_0/a$ and $m_\pi/m_\rho$~\cite{lambda} 
is shown in Fig.~\ref{constant_DIK.ps}.

\section{CRITICAL TEMPERATURE}
\vspace{-1.5mm}
We use non--Abelian and Abelian Polyakov loops to determine the transition
temperature. The Maximally Abelian gauge~\cite{klsw} is fixed by maximizing the
quantity~\cite{bsw}
{\small
\ea
R=\sum\nolimits_{s,\mu}\sum\nolimits_{i=1}^3 |\widetilde{U}_{ii}(s,\mu)|^2\,, \nonumber
\eea
}
with respect to gauge transformations $g$,
$\widetilde{U}(s,\mu)=g(s)U(s,\mu)g^\dagger(s+\hat{\mu})$. We use an $SU(3)$
version of the simulated annealing algorithm~\cite{bbmp}. The Abelian link
variables are defined as $u_i(s,\mu) = \exp\{i\theta^i_{\mu}(s)\}$, where
{\small
\ea
\theta^i_{\mu}(s)
=\arg \widetilde{U}_{ii}(s,\mu)
-\frac{1}{3}\sum\nolimits_{j=1}^3 \arg \widetilde{U}_{jj}(s,\mu)|_{{\rm mod}\,2\pi}\,.\nonumber
\eea
}
We then define the Abelian, monopole and photon Polyakov loop operators
as in~\cite{ploop}.

\begin{figure}[thb]
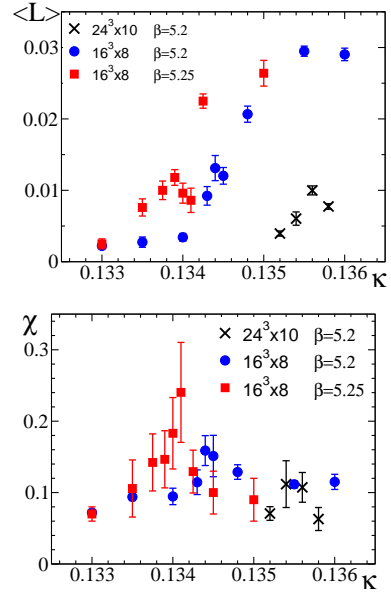

\begin{center}
\includegraphics[angle=0,scale=0.22,clip=true]{polyakov.loop.nonabelian.eps}
\vskip 3mm
\includegraphics[angle=0,scale=0.22,clip=true]{susceptibility.nonabelian.eps}
\end{center}
\vskip -13mm
\caption{\it Non-Abelian Polyakov loop (top) and its susceptibility (bottom).}
\label{NA}
\vskip -3mm
\end{figure}

\begin{figure}[!h]
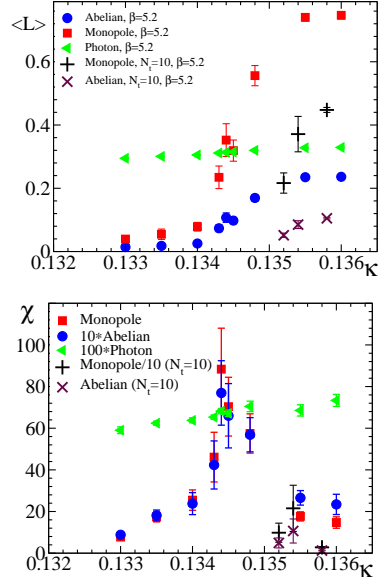

\vskip -1mm
\begin{center}
\includegraphics[angle=0,scale=0.22,clip=true]{polyakov.loop.abelian.eps}
\vskip 3mm
\includegraphics[angle=0,scale=0.22,clip=true]{susceptibility.abelian.eps}
\end{center}
\vskip -11mm
\caption{\it The same as in Fig.~\ref{NA} but for Abelian, monopole and photon contributions.}
\label{AB}
\vskip -8mm
\end{figure}

The Polyakov loop susceptibility is used to determine the transition point.
In Fig.\ref{NA} the non--Abelian Polyakov loop and its susceptibility are
depicted. We get the following values for the critical temperature:
\bc
$T_c \sim 196(8)$MeV,~~~${m_\pi /m_\rho} \sim 0.64$~~{\small\tt(Preliminary)}\\
$T_c=210(4)$MeV,~~~${m_\pi /m_\rho}=0.77$
~~~~~~~~~~~~~~~~~~~\\
$T_c=219(3)$MeV,~~~${m_\pi /m_\rho}=0.81$
~~~~~~~~~~~~~~~~~~~
\ec
It is known~\cite{criexp} that in SU(2) gluodynamics the
non-Abelian, Abelian and monopole Polyakov loops give the
same $T_c$ and critical indices. It follows from comparison of Fig.\ref{NA} and Fig.\ref{AB} that
also in the full QCD these three Polyakov loops have similar
behavior and their susceptibilities have maxima at the same temperature.

\section{CONTINUUM LIMIT}
At small enough lattice spacing and quark mass one can extrapolate the critical
 temperature $T_c$ to the continuum and the chiral limits using formula:
\[
 T_c r_0 = (T_c r_0)^{m_q , a \to 0} +C_a (a/r_0)^2
 +C_q ({1 \over \kappa} - {1 \over \kappa_c})^\alpha\,,
\]
where $r_0 = 0.5$~fm and $(T_c r_0)^{m_q , a \to 0}$ corresponds to the
extrapolated value.

We are brave enough to use four values for $T_c r_0$ (see Table~3), obtained
at rather large quark masses, to estimate the parameters in this extrapolation
expression. A fit gives $(T_c)^{m_q , a \to 0}$ and $\alpha$
with large errors:  $(T_c)^{m_q , a \to 0}\sim 190$~MeV,
$\alpha \sim 0.8$.

{\footnotesize
\bc
\begin{tabular}{|c|c|l|} \hline
$T_c r_0 $&$a/r_0   $&                   \\  \hline
$0.50(2) $&$0.20(1) $&$N_t=10,\beta=5.2$ (prelim.) \\  \hline
$0.53(1) $&$0.234(5)$&$N_t=8,~\beta=5.2$ \\  \hline
$0.55(1) $&$0.225(5)$&$N_t=8,~\beta=5.25$\\  \hline
$0.57(2) $&$0.29(1) $&$N_t=6,~\beta=5.2$\,\, (Ref.~\cite{eh}) \\  \hline
\end{tabular}\\
\vskip 2mm
{\normalsize{Table 3:{\it Available data for $T_c r_0$.}}}
\ec}
In $N_f=2$ QCD the critical indices are expected to belong to the
universality class of the $3D$ $O(4)$ spin model for which one
expects $\alpha=0.55$. Fixing $\alpha$ to this value we get the extrapolated
temperature with a higher accuracy:
\eq\label{Tc} T_c^{m_q , a \to 0}\sim 172.5(3.3)~\mbox{MeV}\,.\ee
This value agrees with values obtained in Refs.\cite{kpe,aakcp}.

\section{CONCLUSIONS}
\begin{figure}[!htb]
\centerline{\includegraphics[angle=0,scale=0.25,clip=true]{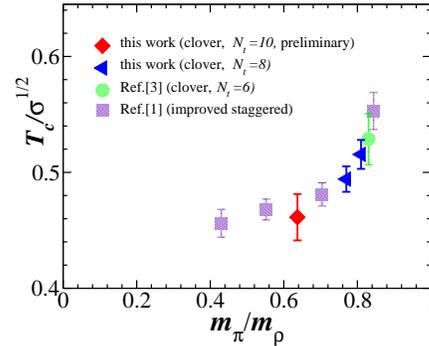}}
\vskip -11mm
\caption{\it $T_c/ \sqrt{\sigma}$, as a function of $m_{\pi} / m_{\rho}$.
The circle and squares show results of \cite{eh} and \cite{kpe} respectively. 
The diamond and triangles correspond to our data.}
\label{tes}
\vskip -8mm
\end{figure}

We determined the critical temperature in full QCD on $16^3 \times 8$ and $24^3
\times 10$ lattices at $\beta =5.2$ and $5.25$ with $N_f=2$ clover fermions
using non-Abelian and Abelian Polyakov loop susceptibilities. 
Our results are in agreement with the results of other groups, as it is shown in
Fig.~\ref{tes}. Assuming that the critical indices of the  two flavor QCD
belong to the universality class of the 3D $O(4)$ spin model, we  extrapolate
the critical temperature $T_c$ to the continuum and the chiral limits. The
extrapolation result is given by (\ref{Tc}). We are continuing simulations on
$24^3 \times 10$ lattice to get better precision of $T_c$ value on this lattice.

\section{ACKNOWLEDGEMENTS}
This work is supported by the SR8000 Supercomputer Project of High Energy
Accelerator Research Organization (KEK). A part of numerical measurements has
been done using NEC SX-5 at Research Center for Nuclear Physics (RCNP) of Osaka
University. M.N.Ch. is supported by JSPS Fellowship P01023. T.S. is partially
supported by JSPS Grant-in-Aid for Scientific Research on Priority Areas
No.13135210 and (B) No.15340073. The Moscow group is partially supported by
RFBR  grants 02-02-17308, 01-02-17456, grants INTAS--00-00111, DFG-RFBR 436 RUS
113/739/0, and CRDF awards RPI-2364-MO-02 and MO-011-0.
G.S. would like to thank Kanazawa University for its kind hospitality.



\begin{thebibliography}{9}
\newcommand{\plb}[1]{Phys. Lett. {\bf B#1}\ }
\newcommand{\npb}[1]{Nucl. Phys. {\bf B#1}\ }
\newcommand{\prd}[1]{Phys. Rev. {\bf D#1}\ }
\bibitem{kpe} F. Karsch, A. Peikert, E. Laermann, \npb{605} (2001) 579.
\bibitem{aakcp} A.A. Khan et al., (CP-PACS), \prd{63} (2001) 034502.
\bibitem{eh} R. G. Edwards, U. M. Heller, \plb{462} (1999) 132.
\bibitem{previous} V. Bornyakov et al., hep-lat/0301003,
Nucl. Phys. {\bf B} (Proc.Suppl.) {\bf 119} (2003) 703.
\bibitem{Jansen:1998mx}
K.~Jansen and R.~Sommer  [ALPHA collaboration],
Nucl.\ Phys.\ {\bf B530} (1998) 185.
\bibitem{lambda} S. Booth et al., \plb{519} (2001) 229.
\bibitem{klsw} A.S. Kronfeld, M.L. Laursen, G. Schierholz, U.-J. Wiese, \plb{198} (1987) 516.
\bibitem{bsw} F. Brandstaeter, G. Schierholz, U.-J. Wiese, \plb{272} (1991) 319.
\bibitem{bbmp} G. Bali, V. Bornyakov, M. M\"uller-Preussker, F. Pahl,
Nucl. Phys. {\bf B} (Proc.Suppl.) {\bf 42} (1995) 852.
\bibitem{ploop} T. Suzuki et al., \plb{347} (1995) 375;
[Erratum-ibid.\ {\bf B351} (1995) 603].
\bibitem{criexp} S. Ejiri, S. Kitahara, T. Suzuki, K. Yasuta, \plb{400} (1997) 163.
\end{thebibliography}
\end{document}